\documentclass[nobm,preprint,info,cite]{epl}
\usepackage{mathlett}

\newcommand{\nc}{\newcommand}
\nc{\vrho}{\varrho}
\nc{\fig}[1]{fig.~\ref{fig:#1}}
\nc{\Eq}[1]{eq.~(\ref{eq:#1})}
\nc{\eq}[1]{(\ref{eq:#1})}

\nc{\ie}{\emph{i.e.,} }
\nc{\cf}{\emph{cf.\ }}

\nc\tauQ{q}
\nc\wQ{{Q}}

\title{Shear flow, viscous heating, and entropy balance\\
   from dynamical systems}

\author{
    Tam\'{a}s T\'{e}l%
    \thanks{E-mail: \email{tel@poe.elte.hu}}\inst{1} 
\and
    J\"{u}rgen Vollmer%
    \thanks{E-mail: \email{vollmer@mpip-mainz.mpg.de}}\inst{2,3}, 
\and
    L\'{a}szl\'{o} M\'{a}ty\'{a}s%
    \thanks{E-mail: \email{matyilac@poe.elte.hu}}\inst{1}
}

\institute{
\inst{1} Inst.~Theor.~Physics,
   E\"{o}tv\"{o}s University,
   P.O.~Box 32,
   H-1518 Budapest, Hungary.
\\
\inst{2} Fachbereich Physik,
   Univ.-GH Essen,
   45117 Essen, Germany.
\\
\inst{3} Max Planck Institute for Polymer Research,
   55128 Mainz, Germany.
}

\shortauthor{T. T\'{e}l \etal}
\shorttitle{Shear flow and entropy balance from dynamical systems}


\begin{document}

\maketitle

\begin{abstract}
  A consistent description of a shear flow, the accompanied viscous
  heating, and the associated entropy balance is given in the
  framework of a deterministic dynamical system, where a multibaker
  dynamics drives two fields: the velocity and the temperature
  distributions.  In an appropriate macroscopic limit their
  transport equations go over into the Navier-Stokes and the heat
  conduction equation of viscous flows. 
  The inclusion of an artificial heat sink can stabilize steady states
  with constant temperatures. It mimics a thermostating algorithm
  used in non-equilibrium molecular-dynamics simulations.
\end{abstract}

\section{Introduction}

In the last years, there has been an increasing interest in
modeling transport phenomena by low-dimensional, deterministic
dynamical systems
\cite{history,Focus,Gasp,G,TVB,Gent,VTB97,GasKla98,GD,TG98,MTV99,GDG00,VMT00}.
Multibaker maps
\cite{G,TVB,Gent,VTB97,GasKla98,GD,TG98,MTV99,GDG00,VMT00}
appeared to be the simplest models of this approach.  They
provide an opportunity to derive the equations of non-equilibrium
thermodynamics from an underlying dynamics \emph{without} using
the concept of particles.  The strongly-chaotic mixing properties
of these two-dimensional maps seem to be sufficient to ensure
consistency with the entropy balance equation of thermodynamics
provided that a properly chosen coarse-grained entropy and a
macroscopic limit are taken \cite{VTB97,TG98,MTV99}. Previous
work successfully described the phenomena of diffusion
\cite{G,Gent}, conduction in an external field 
\cite{TVB,VTB97,GD}, chemical reactions \cite{GasKla98}, thermal
conduction \cite{TG98}, and cross effects due to the simultaneous
presence of an external field and heat conduction \cite{MTV99} by
means of multibaker maps. Not only stationary, but also transient
states could be addressed \cite{VTB97,GDG00,MTV99}. Here we add to
this list the phenomenon of shear flows and the accompanying
viscous heating. The interest of this is to clarify how the shear rate
enters the expression for the irreversible entropy production.
After all, by the definition of local equilibrium the macroscopic
flow profile does not appear in the entropy balance.
Thermodynamic averages contain only \emph{deviations} from the
average streaming velocity.

\begin{figure}
\onefigure[width=\textwidth]{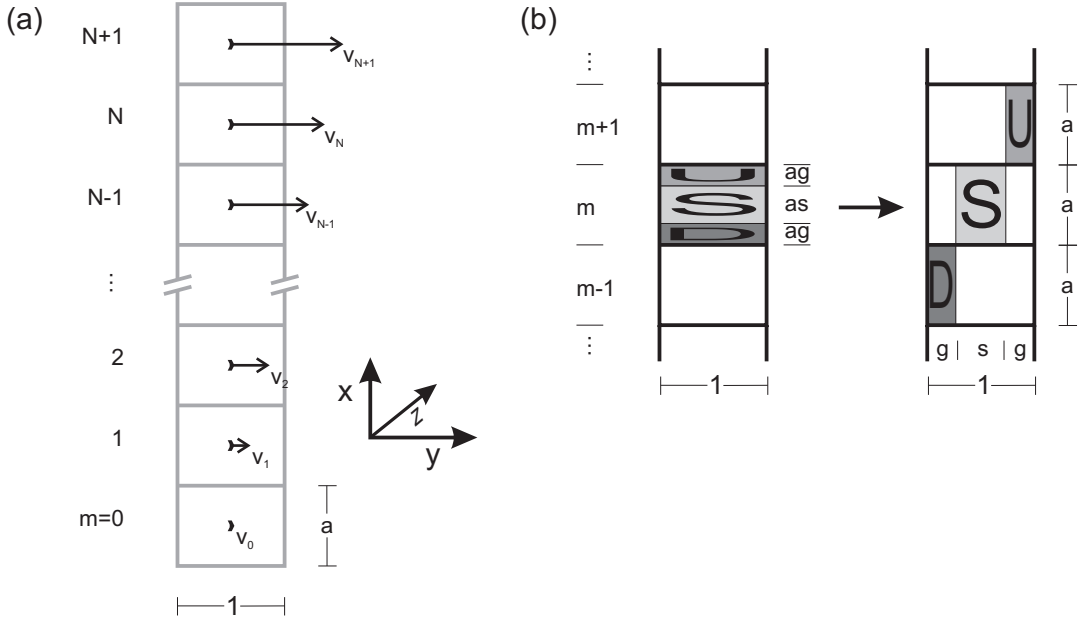} 
%
\caption[]{
The multibaker map. 
a) The chain of $N+2$ baker cells of height $a$
and of unit width modeling an intersection of the fluid along the
$x$ axis. The arrows indicate the average velocity $v_m$ of the fluid
at positions $x\in[am, a(m+1)]$. 
%
b) The 'microscopic' dynamics of cell $m$ ($0<m<N+1$).
All notations are explained in the text. 
}\label{fig:geometry}
\end{figure}

We consider a viscous fluid driven by applying a shear
forcing to two parallel walls of separation $L$
(\cf\cite{GM,B97}). The fluid is assumed to be incompressible.
The upper wall moves in the $y$-direction with a constant velocity
$U$ (fig.~\ref{fig:geometry}a). We restrict ourselves to a
discussion of (possibly non-stationary) 
laminar flow. Due to translation invariance 
the velocity $v$ is always directed in the $y$ direction, and it only
depends on the $x$ coordinate. Similarly, the
temperature $T$ is independent of the $y$ and $z$ coordinates.
There is no external pressure gradient applied, and the internal
pressure forces due to eventual temperature changes are expected
to be negligible. 
For sufficiently long times the well-known linear velocity profile
$v^*(x)=U x/L$ is approached as asymptotic steady state. The spatial
behavior of the temperature distribution $T(x,t)$ depends on the
boundary conditions.  Initially $T$ is typically increasing in time
due to the viscous heating. For an adiabatically closed fluid, the
asymptotic distribution $T^*$ is spatially constant and grows linearly
with time. Inclusion of a heat sink can remove this time dependence
such that the asymptotic state becomes stationary. In a hydrodynamic
setting the sink acts only at the boundaries of the system, but an
artificial, spatially-uniform sink can also be implemented. It mimics
the so-called SLLOD \cite{books} algorithm of non-equilibrium molecular
dynamics.

\section{A multibaker map for shear flow}

We model an intersection of the fluid along the $x$ axis by a
multibaker map.  It consists of $N+2$ cells of size unity times
$a$ (fig.~\ref{fig:geometry}a). The cells $m=0$ and $N+1$
represent the boundary, and cells $m=1,...,N$, the bulk of the
fluid, whose width is $Na=L$. In each cell there is a
momentum-like variable $p$ defined besides the position.  The
multibaker dynamics advects two fields; the 'microscopic' velocity
$v(x,p;t)$ and the 'microscopic' temperature $T(x,p;t)$. The
appropriate (\cf below) averages $v_m$ and $T_m$ over cell $m$
are called the coarse-grained fields. In the macroscopic limit
they go over into the hydrodynamic velocity field and
the thermodynamic temperature, respectively.

The two-dimensional multibaker dynamics is defined as follows
(fig.~\ref{fig:geometry}b). After each time unit $\tau$ every cell
$m=1, \cdots N$ is divided into three bands of heights $ag$, $as$ and
$ag$, such that $g+s+g=1$.  The outermost ones are mapped onto a
column of height $a$ and width $g$ in cells $m+1$ and $m-1$,
respectively. The middle one stays in cell $m$, where it is
transformed into a column of width $s$.
In all cases the area is preserved such that an initially
constant phase-space density $\varrho$ remains constant in time,
reflecting the incompressibility of the fluid. Hence, the density
can be identified with the density of the fluid.

The velocity and the temperature fields are advected by the
multibaker dynamics. In contrast to the density $\varrho$ they
typically evolve in time. Starting with a coarse-grained velocity
distribution $v_m$ along the chain, the updated values $v'_m$
after time $\tau$ become
\begin{equation} \label{eq:vn}
   v'_{m} = v_{m} +g (v_{m-1}+v_{m+1}-2v_m ) ,
\end{equation}
This update expresses momentum conservation in the $x$-direction.
A portion $s=1-2g$ of the original momentum (velocity) remains in
cell $m$, and portions $g$ of the momenta (velocities) of the
neighboring cells flow in.

The temperature equation follows from the energy balance. The full
energy $e_m$ of cell $m$ after coarse graining is proportional to
the sum of the temperature $T_m$ and the translational kinetic
energy $v_m^2/2$, \ie 
\begin{equation}
   e_m = \frac{v_m^2}{2} + C T_m
\label{eq:em}
\end{equation}
where $C$ is a constant. The update of energy is due to an in- and
outflow of energy from the neighbors
\begin{eqnarray}   \label{eq:enprime}
\frac{{(v'_m)}^2}{2} +C T'_m
=
    (1-2g) \left( \frac{v_m^2}{2} + C T_m \right)
+ g \left( \frac{v_{m-1}^2}{2} + C T_{m-1} \right)
+ g \left( \frac{v_{m+1}^2}{2} + C T_{m+1} \right).
\end{eqnarray}
The action of a thermostat can, however, lead to a change of this
kinetic energy. This is modeled by introducing a local source term
$q_m$, and multiplying the energy terms by a factor $[1+\tau
q_m]$, \ie by setting 
$e_m' \rightarrow e_m' \,[1+\tau q_m]$ after every update
(\ref{eq:enprime}).  Rearranging eq.~(\ref{eq:enprime}) subjected to
this additional factor, and using relation \eq{vn} one obtains
\begin{eqnarray}
\label{eq:Tpn}
   T'_m &=&
\biggl\{ T_m + g (T_{m-1}+T_{m+1}-2 T_m) +
   \frac{g}{2 C} [ {(v_{m-1} - v_m)}^2+{(v_{m+1} - v_m)}^2 ] \biggr\}
\; [1+\tau q_m] ,
\end{eqnarray}
which can also be written in the form of the balance equation
\begin{equation}
   \frac{T_m' - T_m}{\tau}
 = Q_m
 - \frac{j_{m+1}^{(w)} - j_m^{(w)}}{a}
\label{eq:Tbal}
\end{equation}
with the discrete ``heat'' current
\begin{mathletters}
\begin{equation}
   j_m^{(w)} = \frac{a^2}{\tau} \;
      g \, \frac{T_m - T_{m-1}}{a} .
\end{equation}
and the source term
\begin{equation}
   Q_m
 = \frac{q_m}{1+\tau q_m} T_m'
  + \frac{g}{2 \tau C} \left[ {(v_{m-1} - v_m)}^2+{(v_{m+1} - v_m)}^2 \right] .
\end{equation}
\end{mathletters}
The first contribution to the source reflects the action of
thermostating, and the second one the effect of viscous heating of
the fluid.

In the absence of coarse graining, a repeated application of the
chaotic, mixing multibaker dynamics leads to a fractal distribution of
the temperature and the velocity fields. Similar to the treatment of
diffusive mass and heat transport \cite{VTB97,MTV99} the emergence and
proliferation of these structures lies at the heart of a consistent
dynamical-systems treatment of static and transient transport
phenomena.

\section{Entropy and entropy balance}

The Gibbs entropy of a multibaker system is defined as the
information-theoretic entropy, \ie as the phase-space average of
$\ln(\varrho/\varrho^*)$. The reference density $\varrho^*$ in the
single-particle phase space of the multibaker map is expected to
depend on the temperature. We take the choice $\varrho^*=\varrho
T^{\gamma}$, where $\gamma$ is a constant exponent. Since the
density $\varrho$ is constant in space and time
the Gibbs energy of cell $m$ becomes%
\footnote{The Boltzmann constant $k_B$ and a constant additive
constant are suppressed here; \cf\cite{VMT00} for details.}
\begin{equation}
   S_m^{(G)}
  =  { \vrho \gamma}
      \int_{\hbox{\tiny cell }m} \upd x\,\upd p \, \ln T(x,p)
\label{eq:SmG}
\end{equation}
The coarse-grained entropy is defined in an analogous way as
\begin{equation}
   S_m=  a \gamma \vrho \ln T_m .
\end{equation}
It is based on the cell-averaged, \ie coarse-grained value of the
temperature $T(x,p)$. As announced, the average (streaming)
velocities $v_m$ do not enter the definition of the entropy.

For the purpose of deriving the time evolution of entropies, it is
useful to choose and initial condition with uniform densities in
every cell. The coarse-grained and the Gibbs entropy then
initially coincide. After one time step, however, the Gibbs
entropy changes due to the fact that the $T$ field takes different
values in the neighboring cells, viz.
\begin{eqnarray}
   S_m^{(G)'}
 &=& {\gamma \vrho} a \;
     \left\{
         g \; \ln[ T_{m-1} \; (1+\tau q_m) ]
       + (1-2g) \; \ln[ T_m \; (1+\tau q_m) ]
       + g \; \ln[ T_{m-1} \; (1+\tau q_m) ]
     \right\}
\nonumber \\
 & = &
    { \vrho \gamma} a \left\{\; \ln[ T_m \; (1+\tau q_m) ]
  + \, g \; \ln\frac{T_{m-1}}{T_m}
  - \, g \; \ln\frac{T_m}{T_{m+1}} \right\} .
\label{eq:SmG'}
\end{eqnarray}
On the other hand, the coarse-grained entropy after one time step
is
\begin{equation}
   S'_m = - {a^2 \gamma \vrho} \ln T_m' .
\label{eq:Sm'}
\end{equation}
Since it only depends on averages in small volumes in the
configuration space, the coarse-grained entropy is considered as
the analogue of the thermodynamic entropy \cite{VTB97,MTV99}. Its
temporal change can be decomposed as
\begin{eqnarray}
  \frac{\Delta S_m}{a \tau}
& \equiv &
   \frac{S_m' - S_m}{a \tau}
=
   \frac{{S_m^{(G)}}' - S_m^{(G)}}{a \tau}
+
   \frac{( {S_m}'- {S_m^{(G)}}' )  -  ( S_m - S_m^{(G)} )}
        {a \tau} ,
\label{eq:micbal}
\end{eqnarray}
and information-theoretic arguments \cite{VTB97,MTV99} lead one to
identify
\begin{mathletters}
\begin{eqnarray}
  \frac{\Delta_e S_m}{a \tau}
&\equiv&
   \frac{{S_m^{(G)}}' - S_m^{(G)}}{a \tau} , 
\label{eq:DefDeSm}
\\
\hbox{and}\qquad 
  \frac{\Delta_i S_m}{a \tau}
&\equiv&
   \frac{( {S_m}'- {S_m^{(G)}}' )  -  ( S_m - S_m^{(G)} )}
        {a \tau}
\label{eq:DefDiSm}
\end{eqnarray}
\end{mathletters}
with the entropy flux and the rate of entropy production,
respectively. Note that the second term of the numerator of
\Eq{DefDiSm} vanishes due to the choice of initial conditions.

Inserting eqs.~\eq{SmG'} and \eq{Sm'} into \Eq{DefDiSm}, 
the rate of entropy production $\Delta_i S_m / (a \tau)$ per
unit volume and time is found to be
\begin{eqnarray} 
   \frac{\Delta_i S_m}{a \tau}
&=&
   \gamma \varrho \tau^{-1}
   \biggl[
      \ln\left( \frac{T_m'}{T_m} (1+\tau q_m)^{-1} \right)
-
   g \ln \frac{T_{m-1}}{T_m} - g \ln \frac{T_{m+1}}{T_m} \biggr] .
\label{eq:DiSm}
\end{eqnarray}
It does not depend on the source term $q_m$, but only on the
values of the field $T$ in cell $m$ and its neighbors. Entropy
production arises in this model from (i) an explicit evolution of
the macroscopic temperature profile, and (ii) from mixing of
regions with different local temperatures. The 
emergence of self-similar structure in the temperature
distribution $T(x,p)$ can lead to a non-vanishing steady-state
entropy production. 
Note that on the level of this discrete relations the entropy 
production does not yet contain the velocity distribution $v_m$. The 
effect of shear flow is only implicitly arising from the splitting 
of the full kinetic energy into a translational and an irregular 
part that depends on the implicit choice of a spatial resolution 
when writing \Eq{em}. 
In the Gibbs entropy the local velocity 
$v(x,p)$ enters implicitly through the definition (\ref{eq:em}) of
the temperature, while for the coarse-grained 
entropy only the averages on the cells (\ie $v_m$) can enter. 

The entropy flux 
$
   {\Delta_e S_m}/{(a \tau)}
$
can be written as the sum of 
the discrete divergence of the entropy current
\begin{mathletters}
\begin{equation}
   j_m^{(s)}
 = - \frac{a \, g}{\tau} \; {\gamma \vrho} \;
   \ln\frac{T_{m+1}}{T_m}
\end{equation}
and the flux
\begin{equation}
   \Phi^{(th)}_m = {\gamma \vrho} q_m
\end{equation}
\end{mathletters}
into the thermostat.

\section{The macroscopic limit}

Establishing a discrete momentum, energy and 
entropy balance is not sufficient to 
motivate the thermodynamic relevance of a dynamical systems model 
transport. As argued in \cite{VTB97} a full consistency can 
only be found in a continuum scaling limit (the \emph{macroscopic 
limit}) where the time evolution equations have 
to coincide with the relations known from irreversible 
thermodynamics, irrespective of the detailed prescription of how to 
choose the discrete time and space units needed to define the 
local equilibrium.  
The macroscopic limit corresponds to  
   $a\ll L$, 
   $N \gg 1$, and 
   $\tau$ 
is much smaller than typical macroscopic time scales (for instance 
the viscous relaxation time). Formally it is defined as 
$
   a,\tau \rightarrow 0
$
such that the spatial coordinate
$
   x = a m
$
is finite. Taking this limit will be indicated by the arrow 
$\rightarrow$. 
 
In order to find the macroscopic limit 
$
   \partial_t v= \nu \, \partial_x^2 v .
$
of the coarse-grained velocity dynamics \Eq{vn} it is required
that $g a^2/\tau$ takes the finite value $\nu$ in the limit, \ie
\begin{equation}
   g \frac{a^2}{\tau} \rightarrow \nu.
\end{equation}
For a vanishing pressure gradient eq.~(\ref{eq:vn}) leads to 
the Navier-Stokes equation for the laminar shear flow \cite{GM}, where
$\nu$ is the \emph{kinematic viscosity}.

Considering the macroscopic limit of eq.~(\ref{eq:Tbal}) one 
obtains the temperature equation 
\begin{equation}  \label{eq:ptTmacr}
\partial_t T
=
\nu \, \partial_x^2 T +
\frac{\nu}{C}  {(\partial_x v_y)}^2
+ T q.
\end{equation}
For $q=0$ this is exactly of the type known from hydrodynamics 
\cite{GM}, where, however, the heat diffusion coefficient 
$\kappa$ appears in front of the second derivative. Hence, in the 
multibaker map $\nu$ also governs heat diffusion, \ie 
$\kappa=\nu$. In the hydrodynamic expression, the coefficient of 
the term expressing viscous heating is $\nu/c_v$, where $c_v$ is 
the specific heat at constant volume. Thus, we have to identify 
the constant $C$ with the specific heat, as also expected from a 
physical interpretation of \Eq{em}.  

Evaluating the macroscopic limit of eq.~(\ref{eq:micbal}) one gets 
for the entropy density $S_m/a \rightarrow s$
\begin{equation}
   \frac{\Delta S_m}{a \tau}
 \rightarrow
   \partial_t s
 = \sigma^{(irr)} - \nabla j^{(s)} +
 \Phi^{(th)}
\label{eq:s-dot}
\end{equation}
which coincides with the thermodynamic entropy balance 
\cite{GM} if $\Phi^{(th)}=0$ in the bulk. For the 
irreversible entropy production we find 
\begin{equation}
   \frac{\Delta_i S_m}{a \tau}
 \rightarrow
   \sigma^{(irr)}
 =
   \gamma \nu \vrho \left( \frac{\partial_x T}{T} \right)^2
 + \frac{\nu \vrho}{T}  \left( \partial_x v \right)^2  .
\end{equation}
It is consistent with thermodynamics if $ \nu \gamma \varrho$ 
corresponds to the heat conductivity $\lambda$ of the flow.  Since 
in general $\lambda = \kappa \varrho c_v$, 
and since $\kappa=\nu$ in our 
case, we conclude that also $\gamma$ amounts to the specific heat in
our model,  
\ie $c_v=C=\gamma$. 

For the entropy current $j_m^{(s)}$ one finds in the macroscopic limit
\begin{equation} 
   j_m^{(s)} 
 \rightarrow
   j^{(s)} (x)
= - \gamma \nu \vrho \frac{\partial_x T}{T}
= - \lambda \frac{\partial_x T}{T} .
\label{eq:jms-th}
\end{equation}
Also this relation fully agrees with its thermodynamic counterpart 
\cite{GM}. 
The heat conductivity $\lambda=\kappa \varrho c_V=\nu \varrho\gamma$
appears in front of the logarithmic derivative ${\partial_x T}/{T}$
without possibility to adjust free parameters, thus demonstrating the
full consistency of the results with irreversible thermodynamics.
Note that eqs.~\eq{s-dot}--\eq{jms-th} are valid at any instant of
time.

\section{Conclusion}

We have established a simple multibaker model that faithfully 
models hydrodynamic and thermal properties of shear flows. 
The model is based on an area-preserving multibaker dynamics. No 
phase-space contraction was needed. Thermostating manifests itself 
in the appearance of the a source term $q$ that only influences 
the heat equation and the entropy flux.  By a proper choice of 
$q(x,t)$ one can stabilize any temperature profile $T^*(x)$ as 
a steady state. For instance the choice 
   $q=\nu  {(\partial_x v)}^2/{(C T)}$ 
eliminates viscous heating, and ensures a $T^*=$constant steady 
state for a uniformly thermostated system.
%
Alternatively, a parabolic stationary temperature profile is 
obtained when the source terms only act at the boundaries in order 
to prescribe a constant temperature at the two ends. 
In either case, the source term $\Phi^{(th)}$ leads to an entropy flux
into a thermostat, and hence requires a generalization of the local
relations of classical irreversible thermodynamics.

The main interest of the model lies in the light it sheds on the
origin of viscous heating in deterministic models of transport. As in
the SLLOD algorithm it arises from the emergence of fractal
structures. In contrast to that model where they are due to the joint
action of a driving force and a Gaussian thermostat introduced by
heuristic arguments into the equations of motion, the present model
also generates the structures for more typical physical settings of
transport driven from the boundaries. It identifies the structures as
arising from the mixing of regions with different \emph{local
  temperatures} that is exponentially proliferating to smaller and
smaller scales for a driven system.  In the present model one can
explicitly follow the analog of an energy cascade in turbulence where
the kinetic energy of a macroscopic flow is distributed to finer and
finer scales until reaching the Kolmogorov scale where it has to be
considered as contributing to the non-directional motion and leads to
viscous heating.  It is exactly this mechanism that also leads to the
appearance of the macroscopic shear rate in the expression of the
irreversible entropy production.

\acknowledgments

We would like to thank Burkhard D\"{u}nweg, Bob Dorfman, Denis Evans, and
Garry Morris for illuminating discussions.
Support from the Hungarian Science Foundation (OTKA Grant No.~032423)
and the Deutsche Forschungsgemeinschaft is acknowledged.




\end{document}